\documentclass[prl,twocolumn,tbtags,superscriptaddress,floatfix]{revtex4}
\usepackage{amssymb}
\usepackage{graphicx,amsmath}
\usepackage{bm}
\usepackage{times}

\begin{document}

\title{Signal amplification in a qubit-resonator system}

\begin{abstract}
We study the dynamics of a qubit-resonator system, when the resonator is
driven by two signals. The interaction of the qubit with the high-amplitude
driving we consider in terms of the qubit dressed states. Interaction of the
dressed qubit with the second probing signal can essentially change the
amplitude of this signal. We calculate the transmission amplitude of the
probe signal through the resonator as a function of the qubit's energy and
the driving frequency detuning. The regions of increase and attenuation of
the transmitted signal are calculated and demonstrated graphically. We
present the influence of the signal parameters on the value of the
amplification, and discuss the values of the qubit-resonator system
parameters for an optimal amplification and attenuation of the weak probe
signal.
\end{abstract}

\date{\today }
\author{D. S. Karpov}
\email{karpov@ilt.kharkov.ua}
\affiliation{B. Verkin Institute for Low Temperature Physics and Engineering, Kharkov,
Ukraine}
\author{G. Oelsner}
\affiliation{Institute of Photonic Technology, Jena, Germany}
\author{S. N. Shevchenko}
\affiliation{B. Verkin Institute for Low Temperature Physics and Engineering, Kharkov,
Ukraine}
\affiliation{V. N. Karazin National University, Kharkov, Ukraine}
\author{Ya. S. Greenberg}
\affiliation{Novosibirsk State Technical University, Novosibirsk, Russia}
\author{E.~Il'ichev}
\affiliation{Institute of Photonic Technology, Jena, Germany}
\maketitle



\section{Introduction}

Quantum optical effects with Josephson-junction-based circuits
have been intensively studied for the last decade. In particular,
such systems are interesting as two-level artificial atoms
(qubits) \cite{OmelIlShev, Wendin,Greenberg08}. Quantum energy
levels and quantum coherence are inherent to qubits and provide
the basis for studying fundamental quantum phenomena. It is
important to note that qubits can be controlled over a wide range
of parameters \cite{Oelsner10, Ashhab09, Wang14, Andersen14,
OmelIlShev} and they have unavoidable coupling to the dissipative
environment.

The ability of stimulated emission and lasing in superconductive devices has
been actively studied during the last several years both theoretically \cite%
{Shevchenko14, Sajko14, Sajko141, Sajko142, Xu14} and experimentally \cite%
{Oelsner13, Hauss08, Forster15}. The work is underway on using
these phenomena as basis for a quantum amplifier of signals near
the quantum limit. This paper was motivated by several recent
publications where the amplification of the input signal was
observed in systems with nanomechanical resonators \cite{Wang14,
Hauss07}, with waveguide resonators \cite{Ashhab09, Hauss08,
Oelsner13, Satanin14,Greenberg07} and the concept of the
amplifiers was discussed \cite{Abdo14, Astafiev101, Lin13,
Bergeal10}.

A key value of the qubit-resonator system in the experiment is the
transmission coefficient of the signal through the resonator. This
transmission coefficient depends on different parameters. The speed and
direction of the energy exchange is determined by relaxation rates. The
variation of the coupling strength allows to change the width of resonance.
The change of the driving amplitude and the magnetic flux (for flux and PSQ
qubit; for charge qubit this quantity is the applied voltage) allows to find
an acceptable point on the resonance line comparative to other parameters.In
the paper we consider how the amplification and attenuation of the input
signal depend on the parameters of the system. The general idea is to find
values and there relationship for the parameters of the system in order to
make the amplification maximal.

In addition to Ref.~\cite{Shevchenko14} here we systematically study the
impact of such parameters as coherent time, resonator loss, coupling and
other. Also we demonstrated how temperature influences the transmission
coefficient. Besides we show the universality of the doubly-dressed approach
for two-level systems. We compare the appearance of the
amplification-attenuation phenomena in both flux and phase-slip qubit \cite%
{Mooij06}

The paper is organized as follows Sec. II contains a description of the
studied system which is a qubit coupled to the two-mode $\lambda /2$
waveguide resonator. Sec. III is devoted to the evolution of the
qubit-resonator system which is described by a Lindblad equation. We analyze
the solution of the Lindblad equation in Sec. IV. Sec. V concludes the paper.

\section{The qubit-resonator system}

The studied system consists of a quantum resonator (transmission-line
resonator with the length $L=\lambda /2$) and a two-level system, the
superconducting flux qubit. The qubit interacts with two harmonics in the
resonator: first probing signal with frequency $\omega _{\mathrm{p}}$ close
to the first harmonic of the resonator and the second signal is a high
amplitude driving signal with frequency $\omega _{\mathrm{d}}$ close to the
third harmonic of the resonator. Such system is analogous to the one studied
recently experimentally in Refs. \cite{Oelsner10, Astafiev12, Shevchenko14}.

\begin{figure}[tph]
\includegraphics[width=8 cm]{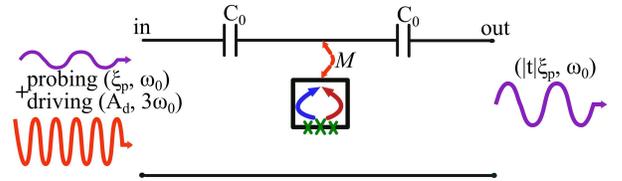}
\caption{Diagrammatic representation of the system under study: a qubit
placed in a waveguide resonator. The qubit interacts with a two-mode
resonator. The first signal has an amplitude $\protect\xi _{\mathrm{d}}$ and
a frequency $\protect\omega _{\mathrm{p}}$. The second signal has an
amplitude $A_{\mathrm{d}}$ and a frequency $\protect\omega _{\mathrm{d}}$. A
measurable (probing) signal at the output has an amplitude different from
the input values.}
\label{Fig: 1}
\end{figure}

\begin{figure}[tph]
\includegraphics[width=8 cm]{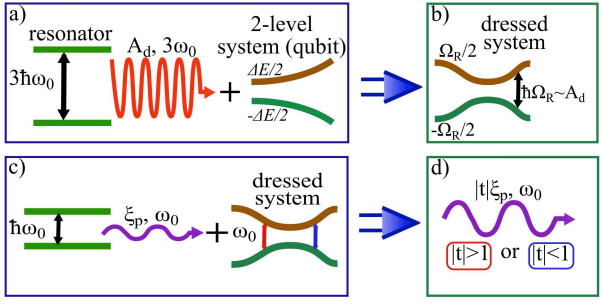}
\caption{The interaction between a qubit and a resonator can be described in
frame of the dressed states. (a) A high-amplitude signal $A_{\mathrm{d}}$
interacts with a two-level system (qubit). (b) Energy levels are modified.
The qubit can be described in terms of the dressed states, in other words we
obtain a dressed two-level system with energy distance proportional to the
amplitude $A_{\mathrm{d}}$ of the third-harmonic. (c) The dressed qubit
interacts with probe signal. (d) The amplitude of the output signal is
increased or weakened.}
\label{Fig2}
\end{figure}

The qubit located in the middle of the resonator ($x=0$) is coupled only to
the odd harmonics $m$, for which the current is defined by $I(x)=I_{{m}}\cos
\pi mx/L$ (see Fig.~\ref{Fig: 1}). The transmission-line resonator runs from
$-L/2$ to $L/2.$ We consider the interaction of the qubit and two-mode
resonator according to doubly-dressed approach, as in Ref.~\cite%
{Shevchenko14}. Hamiltomian of the system is

\begin{eqnarray}
H_{\mathrm{tot}} &=&\hbar \frac{\delta \widetilde{\omega }_{\mathrm{qb}}}{2}%
\widetilde{\sigma }_{z}+\hbar \delta \omega _{\mathrm{r}}a^{\dag }a+
\label{Htot} \\
&&+\hbar \widetilde{\mathrm{g}}\left( a\widetilde{\sigma }^{\dag
}\!+\!a^{\dag }\widetilde{\sigma }\right) +\xi _{\mathrm{p}}\!\left(
a\!+\!a^{\dag }\right) ,  \notag
\end{eqnarray}%
where $\widetilde{\sigma }_{z},$ $\widetilde{\sigma }_{x},$ $\widetilde{%
\sigma }_{y}$, $\widetilde{\sigma }=\frac{1}{2}(\widetilde{\sigma }_{x}-i%
\widetilde{\sigma }_{y})$ are the Pauli's operators in the doubly-dressed
basis; $\delta \widetilde{\omega }_{\mathrm{qb}}=\widetilde{\Delta E}/\hbar
-\omega _{\mathrm{p}}$ \ is the detuning of the doubly-dressed qubit; $%
\widetilde{\Delta E}=\sqrt{\tilde{\varepsilon}^{2}+\tilde{\Delta}^{2}}=$ $%
\hbar \Omega _{R}$ is the Radi frequency of the dressed qubit; $\widetilde{%
\mathrm{g}}=\mathrm{g}_{1}\frac{\varepsilon _{0}}{\Delta E}\frac{\widetilde{%
\Delta }}{\widetilde{\Delta E}}$ is the renormalized coupling; $\Delta E=$ $%
\sqrt{\varepsilon _{0}^{2}+\Delta ^{2}}$; 
$\varepsilon _{0}=2I_{p}\Phi _{0}\left( \Phi _{x}/\Phi _{0}-0.5\right) $%
where $\Phi _{x}$ is the external magnetic flux applied to the qubit loop; $%
I_{\mathrm{p}}$ is the persistent current in the qubit loop; $\Phi _{0}$ is
the flux quantum; $\ \Delta $is the energy separation between two levels at
the degeneracy point $\varepsilon _{0}=0$; $\delta \omega _{\mathrm{r}%
}=\omega _{\mathrm{r}}-\omega _{\mathrm{p}}$ is the detuning of the
resonator; $A_{\mathrm{d}}=4\sqrt{\langle N\rangle }\hbar \mathrm{g}_{%
\mathrm{3}}$ is the normalized amplitude of the driving signal, given by the
average number of photons $N$ in resonator of the third harmonic. The
dressed bias $\widetilde{\varepsilon }$ and the tunneling amplitude $%
\widetilde{\Delta }$ are defined by the driving frequency $\omega _{\mathrm{d%
}}\ $and amplitude $A_{\mathrm{d}}$ either in the weak-driving regime, at $%
A_{\mathrm{d}}<\hbar \omega _{\mathrm{d}}$,%
\begin{equation}
\widetilde{\varepsilon }=\Delta E-\hbar \omega _{\mathrm{d}},\widetilde{%
\Delta }=\Delta A_{\mathrm{d}}/2\Delta E,
\end{equation}%
or in the strong-driving regime, where the energy bias is defined by the
detuning from the $k$-photon resonance, $\widetilde{\varepsilon }\rightarrow
\widetilde{\varepsilon }^{(k)}$, and the renormalized tunneling amplitude is
defined by the oscillating Bessel function, $\widetilde{\Delta }\rightarrow
\widetilde{\Delta }^{(k)}$, as following%
\begin{equation}
\widetilde{\varepsilon }^{(k)}=\Delta E-k\hbar \omega _{\mathrm{d}},\text{ }%
\widetilde{\Delta }^{(k)}=\Delta \frac{k\hbar \omega _{\mathrm{d}}}{%
\varepsilon _{0}}J_{k}\left( \frac{A_{\mathrm{d}}}{\hbar \omega _{\mathrm{d}}%
}\frac{\varepsilon _{0}}{\Delta E}\right) .
\end{equation}

In the doubly-dressed representation the Hamiltonian (\ref{Htot}) is written
for the energy states $|0\rangle $ and $|1\rangle $, where we omitted
constants terms; we have used the rotating-wave approximation. In Fig. \ref%
{Fig2} we explain the processes which take place in the qubit-resonator
system.

\section{Evolution of the system}

One possible method to describe the evolution of an open system is a
solution of the Lindblad equation. In our case we rewrite it in the dressed
basis similar to Ref. \cite{Shevchenko14} and take into account finite
temperature \cite{Scully97}:

\begin{eqnarray}
\frac{d}{dt}\widetilde{\rho } &=&-\frac{i}{\hbar }[H_{\mathrm{tot}},%
\widetilde{\rho }]+\sum\limits_{i}\widetilde{\Lambda }_{i}[\widetilde{\rho }%
],  \label{dpdt} \\
\widetilde{\Lambda }_{\downarrow }[\widetilde{\rho }] &=&\widetilde{\Gamma }%
_{\downarrow }\left( \widetilde{\sigma }\widetilde{\rho }\widetilde{\sigma }%
^{\dag }-\frac{1}{2}\left\{ \widetilde{\sigma }^{\dag }\widetilde{\sigma },%
\widetilde{\rho }\right\} \right) ,  \label{L_down} \\
\widetilde{\Lambda }_{\uparrow }[\widetilde{\rho }] &=&\widetilde{\Gamma }%
_{_{\uparrow }}\left( \widetilde{\sigma }^{\dag }\widetilde{\rho }\widetilde{%
\sigma }-\frac{1}{2}\left\{ \widetilde{\sigma }\widetilde{\sigma }^{\dag },%
\widetilde{\rho }\right\} \right) ,  \label{L_up} \\
\widetilde{\Lambda }_{\phi }[\widetilde{\rho }] &=&\frac{\Gamma _{\phi }}{2}%
\left( \widetilde{\sigma }_{z}\widetilde{\rho }\widetilde{\sigma }_{z}-%
\widetilde{\rho }\right) ,  \label{L_fi} \\
\widetilde{\Lambda }_{\kappa }[\widetilde{\rho }] &=&\kappa \left( \bar{n}%
_{th}+1\right) \left( a\tilde{\rho}a^{\dag }-\frac{1}{2}\left\{ a^{\dag }a,%
\tilde{\rho}\right\} \right) + \\
&&+\kappa \bar{n}_{th}\left( a^{\dag }\tilde{\rho}a-\frac{1}{2}\left\{
aa^{\dag },\tilde{\rho}\right\} \right) ,  \notag \\
\widetilde{\Gamma }_{\downarrow } &=&\Gamma _{1}\bar{n}_{th}\frac{1}{4}%
\left( 1-\frac{\tilde{\varepsilon}}{\widetilde{\Delta E}}\right) ^{2}+ \\
&&+\Gamma _{1}\left( \bar{n}_{th}+1\right) \frac{1}{4}\left( 1+\frac{\tilde{%
\varepsilon}}{\widetilde{\Delta E}}\right) ^{2}+\frac{\Gamma _{\phi }}{2}%
\left( \frac{\tilde{\Delta}}{\widetilde{\Delta E}}\right) ^{2},  \notag \\
\widetilde{\Gamma }_{\uparrow } &=&\Gamma _{1}\bar{n}_{th}\frac{1}{4}\left(
1+\frac{\tilde{\varepsilon}}{\widetilde{\Delta E}}\right) ^{2}+ \\
&&+\Gamma _{1}\left( \bar{n}_{th}+1\right) \frac{1}{4}\left( 1-\frac{\tilde{%
\varepsilon}}{\widetilde{\Delta E}}\right) ^{2}+\frac{\Gamma _{\phi }}{2}%
\left( \frac{\tilde{\Delta}}{\widetilde{\Delta E}}\right) ^{2},  \notag \\
\widetilde{\Gamma }_{\phi } &=&\Gamma _{1}\left( 2\bar{n}_{th}+1\right)
\frac{1}{2}\left( \frac{\tilde{\Delta}}{\widetilde{\Delta E}}\right)
^{2}+\Gamma _{\phi }\left( \frac{\tilde{\varepsilon}}{\widetilde{\Delta E}}%
\right) ^{2}.
\end{eqnarray}%
where $\bar{n}_{th}^{-1}=\exp [\frac{\hbar \nu _{k}}{k_{B}T}]-1$ is the
thermal photon number in the resonator; $\nu _{k}$ is density frequency
distribution for thermal photons; $k_{B}$ is Boltzmann constant; $T$ is
thermodynamic temperature of the system; $\widetilde{\rho }$ is the density
matrix; $\widetilde{\Lambda }_{\phi }[\widetilde{\rho }]$ is the dressed
phase relaxation of the dressed qubit; $\Gamma _{1}$ and $\Gamma _{\phi }$
are the qubit relaxation and dephasing rates; $\widetilde{\Lambda }%
_{\downarrow }[\widetilde{\rho }]$ is the relaxation from $|0\rangle $\ to $%
|1\rangle $ level; $\widetilde{\Lambda }_{\uparrow }[\widetilde{\rho }]$ is
the excitation from $|1\rangle $\ to $|0\rangle $ level. The analysis of
the\ difference between the rates $\widetilde{\Lambda }_{\uparrow }[%
\widetilde{\rho }]$ and $\widetilde{\Lambda }_{\downarrow }[\widetilde{\rho }%
]$ shows availability of the inverse population in the system (Figs.~\ref%
{WrwrGm}~and~\ref{processes}). The equation of motion for the expectation
value of any quantum operator $A$:

\begin{equation}
\frac{d\langle A\rangle }{dt}=-\frac{i}{\hbar }\langle \lbrack
A,H_{tot}]\rangle +Tr(AH_{tot}\sum\limits_{i}\widetilde{\Lambda _{i}}[%
\widetilde{\rho }]),  \label{<A>}
\end{equation}%
where $\langle A\rangle =Tr(A\rho )$, $\langle \lbrack A,H]\rangle
=Tr([A,H]\rho )$, the trace is over all eigenstates of the system; and here $%
H$ is the Hamiltonian of the system in the doubly-dressed basis Eq.~(\ref%
{Htot}). For the expectation values of the operators $a,$ $a^{\dag },$ $%
\widetilde{\sigma }^{\dag },$ $\widetilde{\sigma },$ and $\widetilde{\sigma }%
_{z}$ we obtain the so-called Maxwell-Bloch equations:

\begin{eqnarray}
\frac{d\left\langle a\right\rangle }{dt} &=&-i\delta \omega _{\mathrm{r}%
}^{\prime }\left\langle a\right\rangle -i\widetilde{\mathrm{g}}\left\langle
\widetilde{\sigma }\right\rangle -i\frac{\xi _{\mathrm{p}}}{\hbar },
\label{d<a>/dt} \\
\frac{d\left\langle \widetilde{\sigma }\right\rangle }{dt} &=&-i\delta
\widetilde{\omega }_{\mathrm{qb}}^{\prime }\left\langle \widetilde{\sigma }%
\right\rangle +i\widetilde{\mathrm{g}}\left\langle a\widetilde{\sigma }%
_{z}\right\rangle ,  \label{d<S>/dt} \\
\frac{d\left\langle \widetilde{\sigma }_{z}\right\rangle }{dt} &=&-i2%
\widetilde{\mathrm{g}}\left( \left\langle a\widetilde{\sigma }^{\dag
}\right\rangle -\left\langle a^{\dag }\widetilde{\sigma }\right\rangle
\right) -  \label{dSz/dt} \\
&&-\widetilde{\Gamma }_{+}\left\langle \widetilde{\sigma }_{z}\right\rangle -%
\widetilde{\Gamma }_{-},  \notag
\end{eqnarray}%
where
\begin{eqnarray}
\widetilde{\Gamma }_{\pm } &=&\left( \widetilde{\Gamma }_{\downarrow }\pm
\widetilde{\Gamma }_{\uparrow }\right) , \\
\delta \omega _{\mathrm{r}}^{\prime } &=&\delta \omega _{\mathrm{r}}-i\kappa
/2, \\
\delta \widetilde{\omega }_{\mathrm{qb}}^{\prime } &=&\delta \widetilde{%
\omega }_{\mathrm{qb}}-i\widetilde{\Gamma }_{2}.
\end{eqnarray}%
The Eqs.~(\ref{d<a>/dt})-(\ref{dSz/dt}) were solved in our previous work
\cite{Shevchenko14} in the small photon number limit ($\langle n\rangle \ll
1 $). \ In general, the system of equations is infinite, but it may be
factorized $\langle aa^{\dag }\rangle =\left\langle a\right\rangle
\left\langle a^{\dag }\right\rangle $ etc. This approximation can be used in
the limit of strong perturbation, when the average number of photons in the
system is substantially greater than unity ($\langle n\rangle \gg 1$). In
this way, we simplify Eqs.~(\ref{d<a>/dt})-(\ref{dSz/dt}):

\begin{eqnarray}
\left\langle a\right\rangle &=&-\frac{\xi _{\mathrm{p}}}{\hbar }\frac{\delta
\widetilde{\omega }_{\mathrm{qb}}^{\prime }}{\left\langle \widetilde{\sigma }%
_{z}\right\rangle \widetilde{\mathrm{g}}^{2}+\delta \widetilde{\omega }_{%
\mathrm{qb}}^{\prime }\delta \omega _{\mathrm{r}}^{\prime }},  \label{<a>} \\
\left\langle a^{\dag }\right\rangle &=&-\frac{\xi _{\mathrm{p}}}{\hbar }%
\frac{\delta \widetilde{\omega }_{\mathrm{qb}}^{\prime \ast }}{\left\langle
\widetilde{\sigma }_{z}\right\rangle \widetilde{\mathrm{g}}^{2}+\delta
\widetilde{\omega }_{\mathrm{qb}}^{\prime \ast }\delta \omega _{\mathrm{r}%
}^{\prime \ast }},  \label{<a1>} \\
\widetilde{\Gamma }_{+}\left\langle \widetilde{\sigma }_{z}\right\rangle +%
\widetilde{\Gamma }_{-} &=&2i\frac{\xi _{\mathrm{p}}}{\hbar }\left(
\left\langle a\right\rangle -\left\langle a^{\dag }\right\rangle \right)
+2\kappa \left\langle a\right\rangle \left\langle a^{\dag }\right\rangle .
\label{<Sz>}
\end{eqnarray}

The transmission amplitude of the signal is defined by the formula \cite%
{Scully97, Bishop09, Koshino13}

\begin{equation}
|t|=\frac{\hbar \kappa }{2\xi _{\mathrm{p}}}|\langle a\rangle |.  \label{|t|}
\end{equation}%
The dynamics of the two-level system coupled to a two-mode quantum resonator
can be described as the solution of the Eqs.~(\ref{<a>})-(\ref{<Sz>}) in the
limit of large photon numbers in the resonator. Such description offers a
satisfactory explanation of the experiments with different qubits \cite%
{Oelsner10, Astafiev12}. This is demonstrated below.

\section{Amplification and attenuation of the probe signal}

Consider Eqs.~(\ref{<a>})-(\ref{<Sz>}) in the limit of the weak probing
signal $\xi _{\mathrm{p}}$. We obtain the asymptotic solution for $\langle
\sigma _{z}\rangle $:

\begin{equation}
\langle \sigma _{z}\rangle _{0}=-\frac{\widetilde{\Gamma }_{-}}{\widetilde{%
\Gamma }_{+}}.  \label{<Sz>0}
\end{equation}%
A solution can also be found in the limit of large amplitudes of the probing
signal $\xi _{\mathrm{p}}$:

\begin{equation}
\langle \sigma _{z}\rangle _{\infty }=-\frac{\kappa \left( \widetilde{\Gamma
}_{2}^{2}+\delta \widetilde{\omega }_{\mathrm{qb}}\right) }{\widetilde{%
\Gamma }_{2}\widetilde{\mathrm{g}}^{2}},  \label{<Sz>8}
\end{equation}
where $\widetilde{\Gamma}_{2} = \widetilde{\Gamma}_{\phi}+\frac{\widetilde{%
\Gamma }_{\downarrow }+\widetilde{\Gamma }_{\uparrow }}{2}$.

The Eqs.~(\ref{<a>})~and~(\ref{<Sz>}) were solved in the limit of large
photon numbers in the system $($ $A\mathrm{_{d}}\gg \Delta )$. We obtain two
extremes of the transmission coefficient: amplification (the driving signal
energy is transferred to the probing signal) and attenuation (here vice
versa the probing signal energy is pumped to the driving signal). Consider
the case of full reflection of the probing signal, $|t|=0$. Then Eq.~(\ref%
{<a>}) is simplified
\begin{equation}
\widetilde{\Gamma }_{\phi }+\frac{\widetilde{\Gamma }_{\downarrow }+%
\widetilde{\Gamma }_{\uparrow }}{2}=0.  \label{<a>=0}
\end{equation}%
The left part of Eq.~(\ref{<a>=0}) consists of only positive functions,which
in all experimental parameters space do not come to zero. The full
reflection of the probing signal is impossible. A major effect in studied
system is inverse population. Practically, it is the difference between
excitation and relaxation processes in the dressed qubit,

\begin{equation}
\widetilde{\Gamma }_{\downarrow }-\widetilde{\Gamma }_{\uparrow }=\Gamma _{1}%
\frac{\widetilde{\varepsilon }}{\Delta \widetilde{E}}.  \label{Gn-Gv}
\end{equation}%
The inverse population in the system arises when the relaxation $\widetilde{%
\Gamma }_{-}<0$ or
\begin{equation}
\Delta E<\omega _{\mathrm{d}}.  \label{delE<w_d}
\end{equation}

\begin{figure}[tph]
\includegraphics[width=8cm]{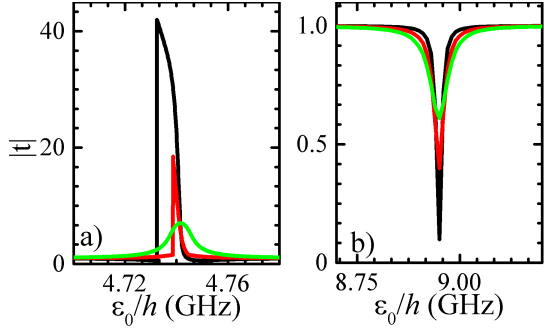}
\caption{ The normalized transmission amplitude as a function of the
normalized magnetic flux $\protect\varepsilon _{0}/h$. We obtain (a) an
amplification and (b) an attenuation of the input signal. The transmission
coefficient strongly depends on relaxations terms. Parameters for (a): $%
\Gamma _{1}/2\protect\pi =4.8$, $6$, $8$~MHz and $\Gamma _{\protect\phi }/2%
\protect\pi =0.15$, $2.8$, $2$~MHz for black, red, and green
curve,respectively; parameters for (b): $\Gamma _{1}/2\protect\pi =0.8$%
\textrm{, }$4.8$, $6$~MHz and $\Gamma _{\protect\phi }/2\protect\pi =0.2$, $%
2 $, $2.8$~MHz for black, red, and green curve, respectively. }
\label{amp_ate}
\end{figure}

Fig.~\ref{amp_ate} is plotted for the following parameters $\Delta /h=3.7$%
~GHz$,$ $\mathrm{g}_{1}/2\pi $ $=0.8$~MHz$,$ $\omega _{\mathrm{r}}/2\pi $ $%
=2.5$~GHz, $\omega _{\mathrm{d}}=3\omega _{\mathrm{r}},$ $\kappa /2\pi =30$%
~kHz, $\xi _{\mathrm{p}}=0.05\kappa $, $\omega _{p}=$ $\omega _{r}$, $A_{%
\text{d}}/h=7$~GHZ. The transmission coefficient sharp changes in the value
at the magnetic flux about $\varepsilon _{0}/h=4.7$~GHz and $\varepsilon
_{0}/h=8.9$~GHz. In the former case, the amplitude of the transmission
signal increases. In the latter case, the transmission signal attenuates.

\begin{figure}[tph]
\includegraphics[width=8cm]{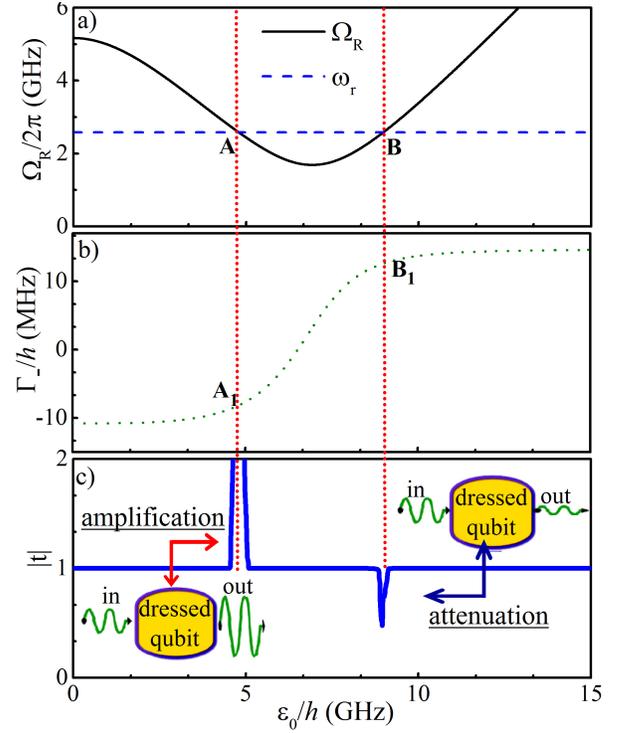}
\caption{A schematic of the processes in the doubly-dressed system. (a) The
black solid line is the Rabi frequency $\Omega _{\mathrm{R}}=\widetilde{%
\Delta E}/h$ of the dressed qubit. The blue dashed line is the resonator
frequency $\protect\omega _{\mathrm{r}}$. In the points of the intersection,
denoted by A and B, the dressed system and the resonator exchange the
energy. The dressed system influences on the passed signal. The output
signal increases or decreases. The type of the process depends on the
population of the energy states (see (b) and (c)). If the relaxation down $%
\widetilde{\Gamma }_{\downarrow }$ is smaller than the excitation $%
\widetilde{\Gamma }_{\uparrow }$, it will be an amplification of the
transmitted signal ($\widetilde{\Gamma }_{-}<0$, point A$_{1}$). When the
opposite situation is realized ($\widetilde{\Gamma }_{-}>0$), it will be an
attenuation. (b) The difference between relaxations from level $|0\rangle $%
to level $|1\rangle $ and from level $|1\rangle $ to level $|0\rangle $.
According to this graphics, we expect an inverse population in the system.
(c) Transmission amplitude as a function of the magnetic flux $\protect%
\varepsilon _{0}.$ }
\label{WrwrGm}
\end{figure}
The amplification of the signal takes place in the system when the Rabi
frequency $\Omega _{\mathrm{R}}$ is close to the resonator frequency (see
Fig.~\ref{WrwrGm}(a),(c)). We obtain the resonant exchange of energy between
the probing signal and the dressed states. The direction of the energy
transfer is specified by the difference between dissipative rates of the
states (see Fig.~\ref{WrwrGm}(b)).

\begin{figure}[tph]
\includegraphics[width=8cm]{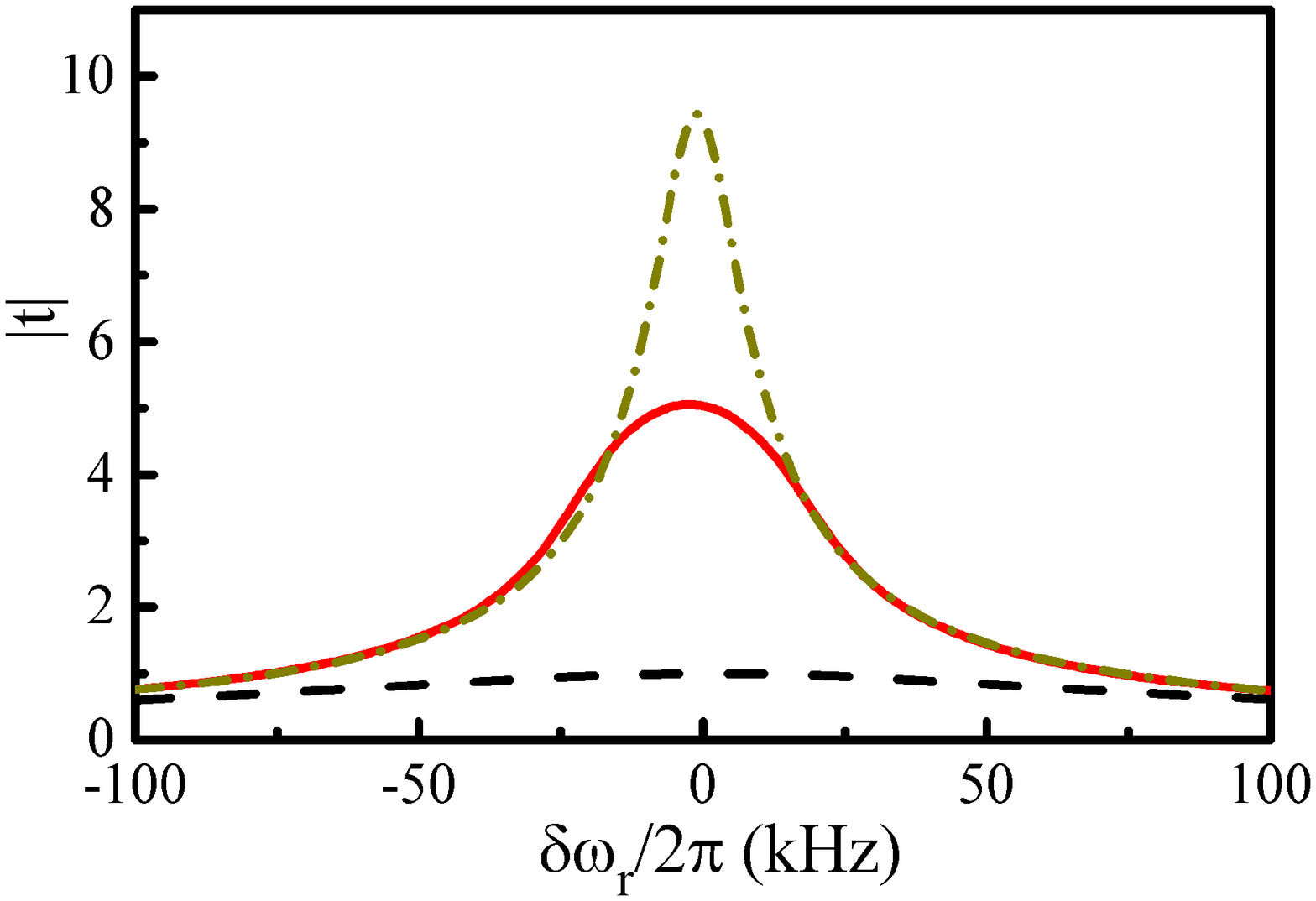}
\caption{Normalized transmission amplitude as a function of the resonator
detuning. \ The dash-dot line and the solid line are plotted for $A_{\text{d}%
}/h=35$~$\mathrm{GHz}$, and the dash line is plotted for $A_{\text{d}}/h=0$.
The parameters of the system are same as for Fig.~\protect\ref{amp_ate}. The
relaxations rates are $\Gamma _{1}/2\protect\pi =9$~MHz$,$ $\Gamma _{\protect%
\phi }/2\protect\pi =4.8$~MHz. The dash-dot and dash lines are plotted for $%
T=0$~K, the solid line is plotted for $T=0.1$~K.}
\end{figure}

\begin{figure}[tph]
\includegraphics[width=8cm]{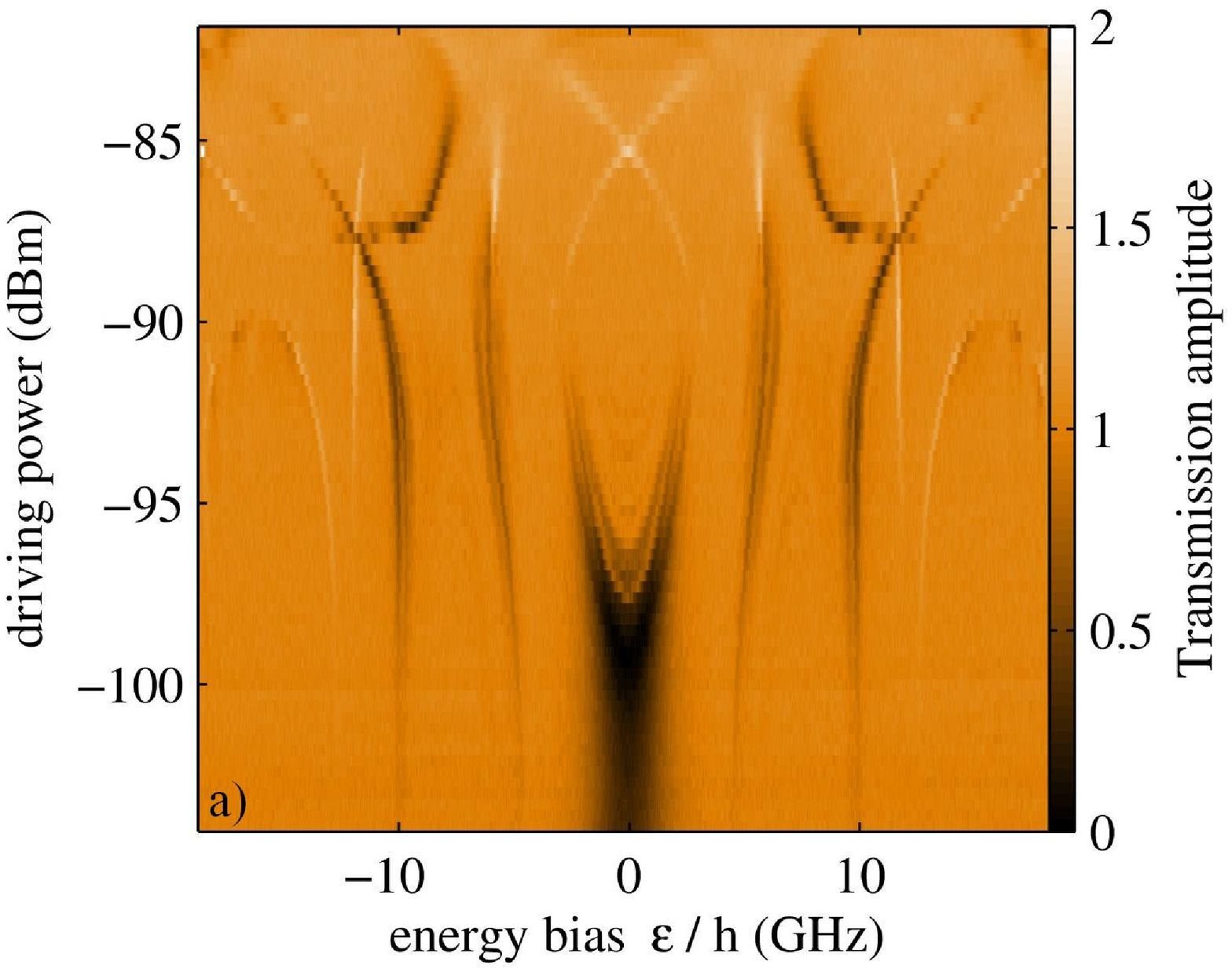}
\includegraphics[width=8cm]{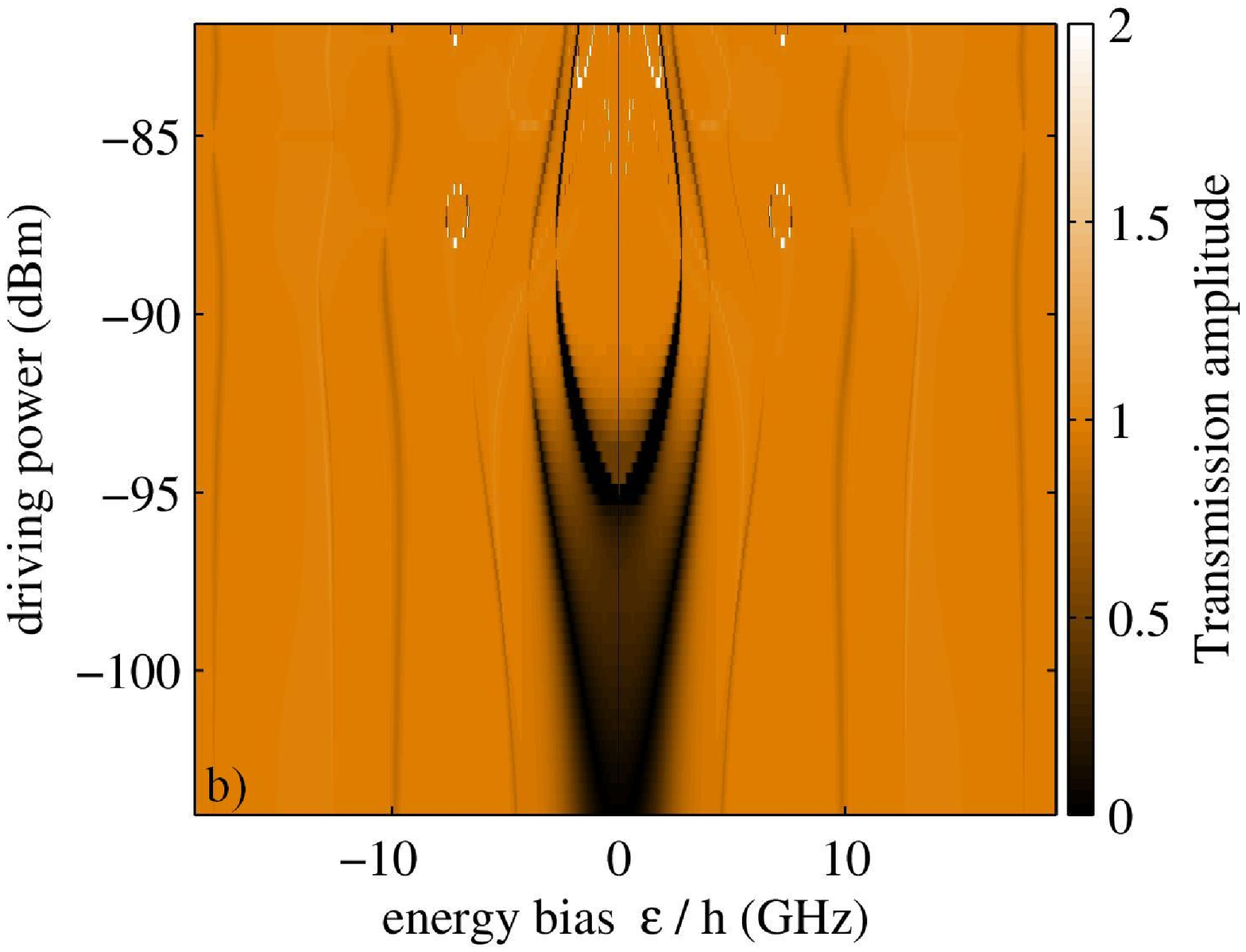} 
\caption{a) Measured normalized transmission amplitude of a probe signal applied at the fundamental mode frequency $\omega_r/2\pi = 2.5$~GHz, while the qubit energy bias $\varepsilon$ and the driving amplitude $A_{\mathrm{d}}$ are varied. The latter is applied in the third harmonic of the resonator. The probing power takes a value of -$127$~dBm. b) Calculation results from (22). The calculation is carried out by splitting the bias axes in parts where the k-th. resonance is dominant but under consideration the energy level shift also induced by the neighboring resonances. The parameters of the qubit- resonator system, $\Delta \approx 3$~GHz, $\mathrm{g}_1/2\pi = 4$~MHz, $\Gamma_{1}/2\pi = 0.75$~MHz, $\Gamma_{\phi}/2\pi = 30$~MHz, and $\kappa/2\pi = 22$~kHz were defined by separate experiments}
\label{t_for_G}
\end{figure}

The analysis of Eqs.~(\ref{<a>})~and~(\ref{<Sz>}) demonstrates that we can
considerably effect on it by varying of the relaxation coefficient and the
amplitudes of the probing and driving signals. In Fig.~\ref{amp_ate} it is
demonstrated how the variation of the relaxation coefficient effects on the
amplification and the attenuation in the system. Such results were
experimentally demonstrated in papers \cite{Oelsner13, Wu77, Khitrova88,
Neilinger15}. The Fig.~\ref{t_for_G} is an example of the variation of the
parameters.

Consider Eqs.~(\ref{<a>})-(\ref{<Sz>}) at the resonance point ($\widetilde{%
\Delta E}=\omega _{\mathrm{p}}$) when the detuning of the resonator $\delta
\omega _{\mathrm{r}}=0$. Then Eqs.~(\ref{<a>})~and~(\ref{<a1>}) are
simplified. We take into account that the average of the operator $\langle
\tilde{\sigma}_{z}\rangle $ under weak probing signal is given by Eq.~(\ref%
{<Sz>0}). For small deviations $\widetilde{\varepsilon }=$ $\Delta E-\hbar
\omega _{d}$ the transmission amplitude is given by the following formula
\begin{equation}
\left\vert \,t\right\vert =\left\vert 1-\frac{8\varepsilon _{0}^{2}\text{g}%
_{1}^{2}\widetilde{\varepsilon }}{\Delta E^{2}Q\widetilde{\Delta }\Gamma
_{1}\kappa }\right\vert ,  \label{|t|simpl}
\end{equation}%
where $Q=\left( (\bar{n}_{th}+\frac{1}{2})+\frac{\Gamma _{\phi }}{\Gamma _{1}%
}\right) \left( \left( 2\bar{n}_{th}+\frac{3}{2}\right) +\frac{\Gamma _{\phi
}}{\Gamma _{1}}\right) .$

The Eq.~(\ref{|t|simpl}) allows to roughly estimate effect of the system
parameters  on the transmission amplitude $t$ as a first approximation. The
non-zero temperature leads to abatement of the amplification. The qubit
relaxation $\Gamma _{1}$ and dephasing $\Gamma _{\phi }$ rates should be
small, then we have system with long coherent time. We can use the
asymptotic of the Bessel function in the limit of the high amplitudes of the
driving signal $A_{\mathrm{d}}$:
\begin{equation}
\widetilde{\Delta }^{(k)}\propto \sqrt{\frac{k^{2}\omega _{\mathrm{d}%
}^{3}\Delta E}{A_{\mathrm{d}}\varepsilon _{0}^{3}}}\propto \frac{1}{\sqrt{A_{%
\mathrm{d}}}}.
\end{equation}%
The transmission coefficient $t$ is proportional to the square root of the
driving amplitude $A_{\mathrm{d}}$ (see Eq.~(\ref{|t|simpl})). The increase
of the driving photon number in the system by preference.

\section{Amplification with phase-slip qubit}

We consider in this section the situation, where there is a so-called
phase-slip qubit (PSQ) coupled to the transmission-line resonator. Our aim
is to clarify similarities and distinctions from the previously considered
case, where we had a flux qubit coupled to the resonator.

\begin{figure}[tph]
\includegraphics[width=8cm]{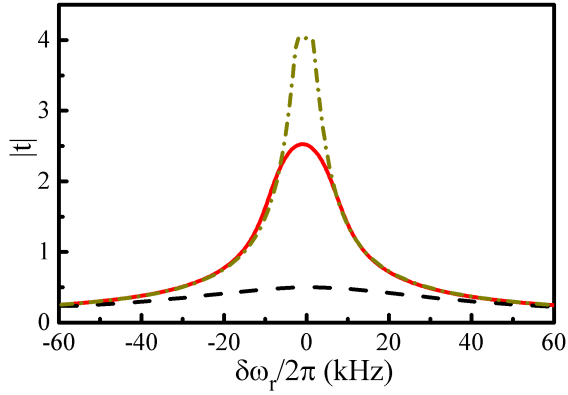}
\caption{ The transmission amplitude for PSQ-resonator system at the value
of the driving amplitude, corresponding to the maximal amplification. The
dash-dot line and the solid line are plotted for $A_{\text{d}}/h=35$~$%
\mathrm{GHz}$, and the dash line is plotted for $A_{\text{d}}/h=0$ .
Parameters for the calculations are the following: $\Delta E/h=$ 4.9~GHz, $%
\protect\omega _{\text{r}}/2\protect\pi =2.4$~GHZ, $\protect\kappa /2\protect%
\pi =$ 30~kHz. Note that the half-width at half-maximum of the transmission
line decreases under pumping. The dash-dot and dash lines is plotted for $%
T=0 $~K, the solid line is plotted for $T=0.1$~K.}
\label{PSQ2}
\end{figure}

\begin{figure}[tph]
\includegraphics[width=8cm]{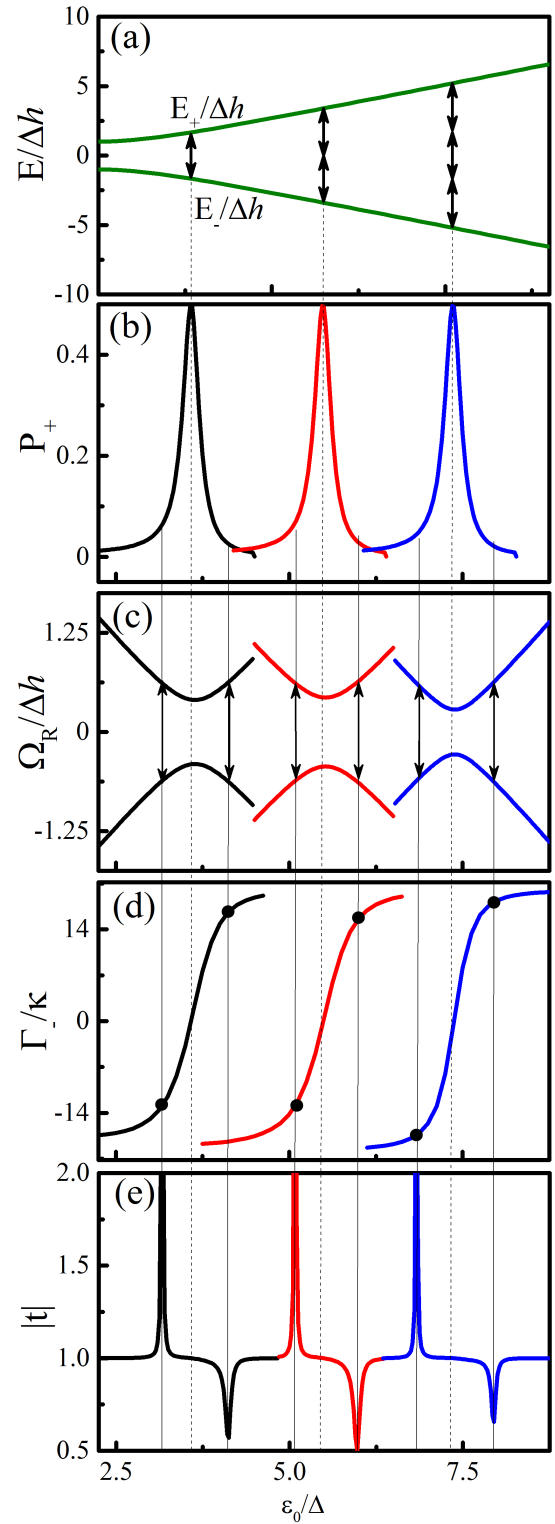}
\caption{ We consider the processes taking place in the qubit-resonator
system. (a) The energy level of the qubit. The one, two and three photon
resonances are marked by the arrows. (b) The maximum value of the excited
state probability of the dressed qubit corresponds to the resonance
condition $\widetilde{\Delta E}=n\hbar \protect\omega _{\mathrm{p}}$ for
integer $n$. (c) The qubit interaction with driving high-amplitude signal
leads to a renormalization of the energy levels of the qubit. (d) The
inverse population is typical for the system with various relaxation times
between the dressed levels. (e) It corresponds to the energy transfer from
the dressed states in the probe signal. When the opposite situation is
realized, the amplitude of the probe signal reduces.}
\label{processes}
\end{figure}

The coherent quantum phase slip has been discussed theoretically in Refs.~%
\cite{Mooij05, Mooij06} and demonstrated experimentally in Ref.~\cite%
{Astafiev12}. It describes a phenomenon exactly dual to the Josephson
effect; whereas the latter is a coherent transfer of charges between
superconducting leads, the former is a coherent transfer of vortices or
fluxes across a superconducting wire. The similar behavior of the coherent
quantum phase slip to Josephson junction allows to consider it as a part of
the qubit-resonator system. The quantum phase slip process is characterized
by the Josephson energy $E_{\mathrm{s}}$, which couples the flux states,
resulting in the Hamiltonian, Ref.~\cite{Mooij05, Mooij06},
\begin{eqnarray}
H &=&-\frac{1}{2}E_{\mathrm{s}}\left( \left\vert N+1\right\rangle
\left\langle N\right\vert +\left\vert N\right\rangle \left\langle
N+1\right\vert \right) + \\
&&+E_{N}\left\vert N\right\rangle \left\langle N\right\vert ,  \notag
\end{eqnarray}%
which is dual to the Hamiltonian of a superconducting island connected to a
reservoir through a Josephson junction; $N$ is the number of the fluxes in
the narrow superconducting wire, $E_{N}=\left( \Phi _{\mathrm{ext}}-N\Phi
_{0}\right) ^{2}/2L_{k}$ is the state energy, $\Phi _{\mathrm{ext}}$ is an
external magnetic flux, $L_{k}$ is the length of the nanowire. The ground
and excited states can be related to the flux basis: $\left\vert
g\right\rangle =\sin \frac{\alpha }{2}\left\vert N\right\rangle +\cos \frac{%
\alpha }{2}\left\vert N+1\right\rangle $ and $\left\vert e\right\rangle
=\cos \frac{\alpha }{2}\left\vert N\right\rangle -\sin \frac{\alpha }{2}%
\left\vert N+1\right\rangle $, where the mixing angle is $\alpha =\arctan E_{%
\mathrm{s}}/\varepsilon _{0}$; the energy splitting between the ground and
excited states is $\Delta E=\sqrt{\varepsilon _{0}^{2}+E_{\mathrm{s}}^{2}}$.
In the rotating wave approximation, the effective Hamiltonian of the system
resonantly driven by a classical microwave field with amplitude $\xi _{%
\mathrm{d}}\cos \left( \Delta E/\hslash \right) $ is $H_{\mathrm{RWA}}=\frac{%
\hslash \Omega }{2}\sigma _{z}.$ Such Hamiltonian coincides with the
Hamiltonian of the flux qubit in the RWA up to the notations \cite{Wendin}.
The interaction between the PSQ and the two-mode resonator can be described
by Hamiltonian~(\ref{Htot}). We demonstrate the transmission coefficient for
a real experimental PSQ in Fig.~(\ref{PSQ2}). We use data which corresponds
to Ref.~\cite{Astafiev12}. Equations~(\ref{<a>})-(\ref{<Sz>}) are also
applicable for the PSQ-resonator system. The doubly-dressed approach is
useful instrument for description of the quantum behavior of the different
mesoscopic systems.

\section{Conclusions}

We studied the evolution of the doubly-driven qubit-resonator
system. We demonstrated the possibility of a large amplification
of the input signal and the ability of almost full reflection of
the probe signal in the system. The value of the transmitted
signal depends on all the system parameters, of
which the coupling coefficient \textrm{g}$_{1}$ and the relaxation rates $%
\kappa $ and $\Gamma _{1,\phi}$ are the most influential. The
numerical simulation of the different qubit-resonator system from
real experimental papers allows to estimate the optimal parameters
for this samples. In particular, we have found that for both
amplification and attenuation the following parameter values are
optimal: \textrm{g}$/\Delta \sim 10^{-4}-10^{-2}$, $\kappa
\lesssim \xi _{p}$, $\Gamma _{1}/\Delta \sim 10^{-2}-10^{-1}$, and
$\Gamma _{\phi}/\Delta \sim 10^{-2}-10^{-1}$. The temperature
noise (non-zero temperature) is diminished the transmission
amplitude.

\begin{acknowledgments}
This work was partly supported by DKNII~(Project No.~M/231-2013),
BMBF~(UKR-2012-028), RFFR~(No.~15-32-50195/15). D.S.K. acknowledges the
hospitality of IPHT (Jena, Germany) and NSTU (Novosibirsk, Russia), where
part of this work was done. The authors are grateful to A.N. Omelyanchouk for useful discussions and comments.
\end{acknowledgments}


\end{document}